\documentclass[superscriptaddress,secnumarabic,
amssymb,amsmath,nobibnotes,aps,prd,showkeys,showpacs,nofootinbib]{revtex4}%
\usepackage{graphicx}
\usepackage{float}
\usepackage{epsf}
\usepackage{bm}
\usepackage{amsmath}
\usepackage{amsfonts}
\usepackage{amssymb}
\usepackage{epstopdf}
\usepackage{natbib}
\usepackage{color}%
\providecommand{\U}[1]{\protect\rule{.1in}{.1in}}
\newcommand{\be}{\begin{equation}}
\newcommand{\ee}{\end{equation}}

\newcommand{\mincir}{\raise
-3.truept\hbox{\rlap{\hbox{$\sim$}}\raise4.truept\hbox{$<$}\ }}
\newcommand{\magcir}{\raise
-3.truept\hbox{\rlap{\hbox{$\sim$}}\raise4.truept\hbox{$>$}\ }}

\begin{document}

\title{{ Reheating constraints in Instant Preheating}}

\author{Jaume de Haro\footnote{E-mail: jaime.haro@upc.edu}}
\affiliation{Departament de Matem\`atiques, Universitat Polit\`ecnica de Catalunya, Diagonal 647, 08028 Barcelona, Spain}


\thispagestyle{empty}

\begin{abstract}

We use Instant Preheating as a mechanism to reheat the universe when its evolution is modeled by a non-oscillating background. Once we obtain the reheating temperature, we calculate the number of e-folds using two different methods, which allows us to establish a relationship between the reheating temperature and the spectral index of scalar perturbations. We explore this connection to constrain the spectral index for different Quintessential Inflation models.

\end{abstract}

\vspace{0.5cm}


\maketitle

\thispagestyle{empty}

\section{Introduction}

The reheating temperature and the spectral index of scalar perturbations are closely linked in inflationary cosmologies. Therefore, by establishing the relationship between them and determining the range of viable reheating temperatures, we can calculate the possible values of the spectral index and compare them with the observational data provided by Planck's team.

\

With this idea in mind, we calculated the number of e-folds for non-oscillating inflationary models in two different ways. First, we used observational data from the present to the end of inflation, which allowed us to determine the number of e-folds as a function of the reheating temperature and the spectral index. Second, we used the inflationary potential, which states that the number of e-folds depends solely on the spectral index. So, 
by equating both expressions, we established the relationship between the reheating temperature and the spectral index.

\

The next step is to investigate the relationship between the reheating temperature and the spectral index when the reheating mechanism is the well-known {\it Instant Preheating} \cite{fkl0,fkl}. We apply our results to various Quintessential Inflation (QI) scenarios, such as the Peebles-Vilenkin model \cite{pv}, exponential $\alpha$-attractors \cite{dimopoulos}, and double exponential models \cite{Geng}.
One of our main findings concerning Instant Preheating is that the coupling constant, denoted by $\tilde{g}$, between the inflaton field and the quantum field responsible for particle production is highly constrained. We found that its value must lie between $10^{-6}$ and $10^{-5}$. The lower limit is necessary to prevent vacuum polarization effects during the last e-folds of inflation from affecting the evolution of the inflaton field. The upper limit is due to the requirement that the reheating temperature is below $10^9$ GeV to avoid interference with the success of Big Bang Nucleosynthesis (BBN) \cite{ellis}, caused by the late decay of gravitationally interacting particles, such as the gravitino or the moduli fields.
For a coupling value around $\tilde{g}\cong 5\times 10^{-6}$, shortly after the start of kination, the created particles become non-relativistic, and during the kination phase, they decay into lighter particles, which reheats the universe to a temperature restricted to the range of $10^{-12} M_{pl}$ to $10^{-10} M_{pl}$, where $M_{pl}$ is the reduced Planck mass.

\

After obtaining the maximum and minimum values of the reheating temperature, we use the link between the reheating temperature and the spectral index of the scalar perturbations for a given Quintessential Inflation (QI) model to constrain it. This results in a narrow range of viable values, which falls within the 2$\sigma$ Confidence Level of the observable values obtained by the Planck's team.

\

Finally, we also investigate the implications of this relationship when reheating occurs via gravitational particle production. Specifically, we consider an exponential $\alpha$-attractor model and show the interconnection between the spectral index and the mass of the produced particles. We find that for a spectral index close to $n_s\cong 0.97$, there are heavy as well as light masses that can give rise to viable reheating temperatures ranging from $1$ MeV to $10^7$ GeV.

\

The present work is organized as follows: In Section II, we investigate the Instant Preheating mechanism and obtain the range of viable reheating temperatures. In Section III, we present the two different methods used for calculating the number of last e-folds and the relationship between the spectral index and the reheating temperature. In Section IV, we apply this relationship to different QI models. In Section V, we consider reheating via gravitational particle production and apply it to $\alpha$-attractors to compute feasible reheating temperatures. Finally, in the concluding Section, we summarize our findings and present our conclusions.

\section{Instant Preheating}\label{instant-preheating}

In this section, we will review one of the most commonly used reheating mechanisms for non-oscillating models, known as Instant Preheating, which was introduced by Felder, Kofman, and Linde in \cite{fkl0,fkl}. The basic concept is that the inflaton field, denoted as $\varphi$, is coupled to a scalar quantum field $\phi$, and this coupling is responsible for particle production.

\

 The Lagrangian density of the quantum field $\phi$ is given by 
\begin{eqnarray}\label{Lagrangian-density}
    {\mathcal L}=\frac{1}{2}\sqrt{|g|}(g^{\mu\nu}\partial_{\mu}\phi
    \partial_{\nu}\phi-(m^2+\tilde{g}^2(\varphi-\varphi_{kin})^2)    \phi^2-\xi R\phi^2),\end{eqnarray}
where $m$ is the bare mass of the field, $R$ is the Ricci scalar, $\varphi_{kin}$ is the value of the inflaton at the beginning of kination, and $\tilde g$ is the dimensionless coupling constant between the inflation field and the quantum field. Considering conformally coupled particles, i.e., choosing $\xi=1/6$, the frequency of the modes will be given by:
\begin{eqnarray}\label{instant0}\omega_k^2(\eta)=k^2+m_{eff}^2(\eta)a^2(\eta),\end{eqnarray}
where $m_{eff}(\eta)=\sqrt{m^2+\tilde{g}^2
(\varphi(\eta)-\varphi_{kin})^2}$ is  the effective mass of the produced particles.

\
 
The analytic computation of the Bogoliubov coefficients is based on the linear  approximation
$\varphi(\eta)-\varphi_{kin}\cong \varphi_{kin}'(\eta-\eta_{kin})$  and the assumption that  the  universe is static with $a(\eta)=a_{kin}$. Then, the frequency becomes 
\begin{eqnarray}\omega_k^2(\eta)=k^2+(m^2+
\tilde{g}^2 (\varphi'_{kin})^2(\eta-\eta_{kin})^2)a_{kin}^2\end{eqnarray}
and,  thus,  the analytic value of the $\beta$-Bogoliubov coefficients is given by \cite{fkl}:
\begin{eqnarray}\label{instant}
    |\beta_k|^2\cong \exp \left({-\frac{\pi (k^2+m^2a^2_{kin})}{\tilde{g}a_{kin}\varphi'_{kin}}}\right)
    = \exp\left({-\frac{\pi (k^2+m^2a^2_{kin})}{\sqrt{6}\tilde{g}a_{kin}^2H_{kin}M_{pl}}}\right).
     \end{eqnarray}

\

This last formula was tested numerically in \cite{campos} for the original Peebles-Vilenkin model \cite{pv}, 
and  also in \cite{haro23} for the non-oscillating background 
\begin{eqnarray}\label{background}
a^2(\eta)=\frac{1}{2}\left[
\left(1-\tanh(\eta/\Delta\eta)\right)\frac{1}{1+H_{inf}^2\eta^2}+
(1+\tanh(\eta/\Delta\eta))(3+2H_{inf}\eta)
\right],    
\end{eqnarray}
where the scale of inflation, denoted by $H_{inf}$, is typically of the order $10^{-6} M_{pl}$ in the majority of inflationary models. The time scale of the phase transition from the end of inflation to the beginning of kination is represented by $\Delta\eta$.

\

Note that the production of particles is exponentially suppressed for large values of the bare mass, due to the form of the $\beta$-Bogoliubov coefficient. Therefore, we will set $m=0$, which yields $m_{eff}=\tilde{g}(\varphi-\varphi_{kin})$. Additionally, the modes that contribute to the particle production are those satisfying
\begin{eqnarray}
    \frac{k^2}{a^2_{kin}}< \tilde{g} H_{kin} M_{pl},
\end{eqnarray}
then, in order to have non-relativistic particles during kination, which $\omega_k(\eta)\cong a(\eta) m_{eff}(\eta)$, we need to  demand 
\begin{eqnarray}
\tilde{g}H_{kin}M_{pl}<\tilde{g}^2(\varphi(\eta)-\varphi_{kin})^2\cong \tilde{g}^2M_{pl}^2
\ln^2\left(\frac{H_{kin}}{H(\eta)} \right),
\end{eqnarray}
where we have used that during kination,
the inflaton field evolves according to:
\begin{eqnarray}\label{inflaton-evolution}
\varphi(\eta)=\varphi_{kin}+\sqrt{\frac{2}{3}}M_{pl}\ln \left( \frac{H_{kin}}{H(\eta)} \right).
\end{eqnarray}

Therefore, 
if we consider $H(\eta) < H_{kin}/3$, then the quantity $\ln \left( \frac{H_{kin}}{H(\eta)} \right)$ is greater than $1$. Additionally, in the majority of inflationary models, we have $H_{kin} \cong 10^{-7}M_{pl}$.
So, by imposing the following condition:
\begin{eqnarray}
\tilde{g}H_{kin}M_{pl} < \tilde{g}^2M_{pl}^2 \Longrightarrow \tilde{g} > H_{kin}/M_{pl} \cong 10^{-7},
\end{eqnarray}
we can ensure that the particles become non-relativistic shortly after the start of kination.

\

After the beginning of kination, when the non-relativistic particles have already been created, the inflaton field evolves according to:
\begin{eqnarray}\label{Xx}
\ddot{\varphi}+3H\dot{\varphi}=-\tilde{g}\langle \hat\phi^2\rangle m_{eff},
\end{eqnarray}
where, $\langle \hat\phi^2\rangle$ is the renormalized vacuum average of the quantum operator $\hat{\phi}^2$. To prevent an undesirable second inflationary period, we need to demand that the right-hand side of Eq. (\ref{Xx}) is subdominant before the decay of these non-relativistic particles. Therefore, before the decay, we need to impose the following condition:
\begin{eqnarray}\label{condition}
H\dot{\varphi} \gg \tilde{g}\langle \hat\phi^2\rangle m_{eff}.
\end{eqnarray}

Effectively, if the right-hand side of Eq. (\ref{Xx}) ceases to be negligible, the inflaton field would be under the action of the quadratic potential given by:
\begin{eqnarray}
V(\varphi) =\frac{m_{eff}^2}{2}\langle \hat{\phi}^2\rangle=\frac{1}{2}\tilde{g}^2(\varphi-\varphi_{kin})^2\langle \hat{\phi}^2\rangle,
\end{eqnarray}

As a result, the field will roll down to $\varphi_{kin}$, which could potentially initiate a new inflationary phase that we do not desire.

\

Therefore, taking into account that for non-relativistic particles the evolution of $\phi$ is approximately that of a harmonic oscillator:
\begin{eqnarray}
(a\phi)''+a^2(\eta)m_{eff}^2(\eta)(a\phi)=0,
\end{eqnarray}
because the term $\Delta(a\phi)$ is negligible for non-relativistic particles, we have that its evolution is like:
\begin{eqnarray}
a\phi \propto e^{-i\int am_{eff}} \Longrightarrow (a\phi)'\sim
a(\eta)m_{eff}(\eta) (a\phi),
\end{eqnarray}
meaning that the re-normalized vacuum energy density, which is the effective mass multiplied by the number density of produced particles, is like that of a harmonic oscillator, i.e.,
\begin{eqnarray}
\langle \hat{\rho}(\eta)\rangle
= m_{eff}(\eta)\langle \hat{N}(\eta)\rangle
\cong \frac{1}{2a^4(\eta)}
\langle ((a\hat{\phi})')^2+ a^2(\eta)m_{eff}^2(\eta) (a\hat{\phi})^2\rangle \cong m_{eff}^2(\eta) \langle \hat{\phi}^2(\eta)\rangle,
\end{eqnarray}
leading to \cite{fkl} (see also the Appendix of this work, where a more rigorous demonstration was done):
\begin{eqnarray}
\langle \hat{\phi}^2(t)\rangle \cong
\frac{ \langle \hat{N}(t)\rangle}{m_{eff}(t)},
\end{eqnarray}
and recalling that during kination $H\sim \dot{\varphi}/M_{pl}$,
the condition (\ref{condition}) becomes:
\begin{eqnarray}
\dot{\varphi}^2(t)\gg \tilde{g}M_{pl} \langle \hat{N}(t)\rangle
\Longrightarrow \rho_B(t)\gg \tilde{g}M_{pl} \langle \hat{N}(t)\rangle,
\end{eqnarray}
where $\rho_B(t)= \frac{\dot{\varphi}^2(t)}{2}$ is the energy density of the background in the kination phase.

\

Shortly after the beginning of kination, as we have already shown, the effective mass of the produced particles becomes greater than $\tilde{g}M_{pl}$, meaning that if they decay into lighter ones to reheat the universe before the end of kination, i.e., if $\rho_B(t)\gg \langle \hat{\rho}(t)\rangle\cong m_{eff}(t)\langle \hat{N}(t)\rangle$ before their decay, the bound (\ref{condition}) will be automatically satisfied. Then, to ensure that the inflaton field rolls towards infinity, as in all QI models, we will assume that the decay of the produced particles into lighter ones occurs before the end of kination.

\

Now, we calculate the energy densities at the time of decay, which occurs when $H\sim \Gamma$, where $\Gamma$ is the decay rate. The corresponding energy densities are
\begin{eqnarray}
\rho_{B,dec}=3\Gamma^2M_{pl}^2\qquad \mbox{and} \qquad
\langle \hat{\rho}_{dec}\rangle\cong m_{dec}\frac{\Gamma}{H_{kin}}\langle \hat{N}_{kin}\rangle,
\end{eqnarray}
where $m_{dec}\equiv m_{eff}(t_{dec})$ and
we have used that during kination the Hubble rate scales as $a^{-3}$, which implies $\left(\frac{a_{kin}}{a_{dec}} \right)^3=\frac{\Gamma}{H_{kin}}$.

\

After the decay,  the energy densities evolve as
\begin{eqnarray}
    \rho_{B}(t)=3\Gamma^2M_{pl}^2\left(\frac{a_{dec}}{a(t)} \right)^6 \qquad \mbox{and} \qquad
    \langle \hat{\rho}(t)\rangle\cong m_{dec}\frac{\Gamma}{H_{kin}}\langle  \hat{N}_{kin}\rangle
    \left(\frac{a_{dec}}{a(t)} \right)^4,\end{eqnarray}
and since the reheating occurs at the end of kination, i.e., when 
$\rho_{B}(t)\sim \langle \hat{\rho}(t)\rangle$,  we have
\begin{eqnarray}
\left(\frac{a_{dec}}{a_{reh}} \right)^2=\frac{m_{dec}\langle  \hat{N}_{kin}\rangle}{ 3\Gamma H_{kin} M_{pl}^2},
\end{eqnarray}
and thus, using the Stefan-Boltzmann law,  the reheating temperature is given by: 
\begin{eqnarray}
    T_{reh}=\left(\frac{30}{\pi^2 g_{reh}} \right)^{1/4}
    \langle \hat{\rho}_{reh}\rangle^{1/4}=
\left(\frac{10}{3\pi^2 g_{reh}}\right)^{1/4}\left(\frac{m_{dec}\langle  \hat{N}_{kin}\rangle}{ \Gamma^{1/3} H_{kin} M_{pl}^{8/3}}\right)^{3/4}M_{pl}\nonumber \\
=\left(\frac{5\sqrt{3}}{\pi^{11} g_{reh}} \right)^{1/4}  
\left(\tilde{g}^{3/2}\frac{m_{dec}\rho_{B,kin}^{1/4}}{ \Gamma^{1/3}  M_{pl}^{5/3}}\right)^{3/4}
M_{pl},
\end{eqnarray}
where $g_{reh}=106.75$ is the effective number of degrees of freedom for the Standard Model,  and we have taken into account that:
\begin{eqnarray}
   \langle \hat{N}_{kin}\rangle=\frac{1}{2\pi^2a_{kin}^3}\int_0^{\infty}
   k^2|\beta_k|^2 dk
   = \frac{1}{8\pi^3}(\tilde{g} \sqrt{2\rho_{B,kin}})^{3/2}.
    \end{eqnarray}

\

 After some algebra, one has 
\begin{eqnarray}
    T_{reh}\cong 2\times 10^{-2}\tilde{g}^{15/8}\left(
\frac{\sqrt{H_{kin}}}{\Gamma^{1/3}M_{pl}^{1/6}}
    \right)^{3/4}\ln^{3/4}\left( \frac{H_{kin}}{\Gamma}\right)M_{pl},
\end{eqnarray}
which for  $H_{kin}\cong 10^{-7}M_{pl}$ becomes
\begin{eqnarray}
    T_{reh}\cong 3\times 10^{-3}\tilde{g}^{15/8}\left(
\frac{M_{pl} }{\bar\Gamma}
    \right)^{1/4}\ln^{3/4}\left( \frac{M_{pl}}{\bar\Gamma}\right)M_{pl},
\end{eqnarray}
where we have introduced the notation $\bar\Gamma\equiv 10^7 \Gamma$.

\

On the other hand, the condition that the decay occurs during kination leads to the constraint:

{
\begin{eqnarray}\label{bound1}
\frac{\sqrt{2}}{3\sqrt{3}}10^{14}\ln\left(\frac{M_{pl}}{\bar\Gamma} \right)
\frac{\tilde{g}\langle \hat{N}_{kin}\rangle}{M_{pl}^2}\leq\bar\Gamma < M_{pl}/3
\Longleftrightarrow
10 \tilde{g}^{5/2}\ln\left(\frac{M_{pl}}{\bar\Gamma} \right)
\leq \frac{\bar\Gamma}{M_{pl}}< \frac{1}{3},
\end{eqnarray}
}
where the condition $\bar\Gamma<M_{pl}/3$ comes from the fact that by imposing it, we have $m_{dec}\geq \tilde{g}M_{pl}$. For $\tilde{g}>10^{-7}$, as we have already shown, this ensures that the decaying particles are non-relativistic.

\

Noticing that a viable reheating temperature should be above $1$ MeV, as this is the temperature at which Big Bang nucleosynthesis (BBN) occurs, we have:
\begin{eqnarray}
    5\times 10^{-22}M_{pl}\leq T_{reh}<\sqrt{M_{pl}\Gamma}
    \Longrightarrow \frac{\bar\Gamma}{M_{pl}}\geq 10^{-36}
    \Longrightarrow \ln\left(\frac{M_{pl}}{\bar\Gamma} \right)\leq 10^2,\end{eqnarray}
where we have used the fact that the decay occurs before reheating. This last restriction implies that the reheating temperature is bounded by
\begin{eqnarray}
    T_{reh}\leq  10^{-1}\tilde{g}^{15/8}\left(
\frac{M_{pl} }{\bar\Gamma}
    \right)^{1/4} M_{pl},\end{eqnarray}
which, in order to ensure that the reheating temperature is bellow $5\times 10^{-10} M_{pl}\cong 10^9$ GeV,  leads to 
\begin{eqnarray}
    \frac{\bar\Gamma}{M_{pl}}
     \geq 2\times 10^{33}\tilde{g}^{15/2}.\end{eqnarray}

{
The condition 
$2\times 10^{33}\tilde{g}^{15/2}\geq 10^3 \tilde{g}^{5/2},$
implies $\tilde{g}\geq 10^{-6}.$}
Consequently, by choosing $10^{-6}\leq \tilde{g}\leq 3\times 10^{-5}$, we ensure that $m_{dec}< M_{pl}$, avoiding problems during BBN. If the effective mass becomes greater than the Planck's mass, each particle would become a Planck-sized black hole, which would immediately evaporate and produce gravitinos or moduli fields. Thus, a late decay could potentially jeopardize the success of BBN. With this condition satisfied, the constraint (\ref{bound1}) becomes:
\begin{eqnarray}
    2\times 10^{33}\tilde{g}^{15/2}
    \leq  \frac{\bar\Gamma}{M_{pl}} <1/3.\end{eqnarray}

\

Finally, it is also important to ensure that the vacuum fluctuations do not disturb the evolution of the inflaton during the last stages of inflation, which is accomplished by imposing that $m_{eff}(t)\geq H(t)$.
Noticing that during the last stage of inflation the effective mass is of the order $\tilde{g}M_{pl}$ and assuming, as in most inflationary models, that the scale of inflation is of the order $10^{-6} M_{pl}$, one has to impose $\tilde{g}\geq 10^{-6}$. We can show this in the case of a quadratic potential $V(\varphi)=\frac{M^2}{2}(\varphi-\varphi_{kin})^2$ with mass $M$, where the power spectrum of scalar perturbations 
\begin{eqnarray}\label{power}
        {\mathcal P}_{\zeta}=\frac{H_*^2}{8\pi^2 M_{pl}^2\epsilon_*}\cong 2\times 10^{-9},
    \end{eqnarray}
(the "star" means that the quantities are evaluated at the horizon crossing), together with the slow roll parameters $\epsilon_*$ and $\eta_*$, and the well-known relation $1-n_s=6\epsilon_*-2\eta_*$, where $n_s$ is the spectral index,  tells us that the value of the mass is $M\sim 16\pi \sqrt{0.3} (1-n_s)10^{-4} M_{pl}$. Then, the condition $m_{eff}(t)\geq H(t)$ becomes:
\begin{eqnarray}
    \tilde{g}^2(\varphi-\varphi_{kin})^2\geq 
    V(\varphi)=\frac{M^2}{6M_{pl}^2}(\varphi-\varphi_{kin})^2   \Longrightarrow
    \tilde{g}\geq \frac{3}{4}(1-n_s)10^{-4}\cong 10^{-6},
    \end{eqnarray}
where we have taken the conservative value $1-n_s\cong 10^{-2}$.

\

Thus, a successful reheating of the universe is achieved through the Instant Preheating mechanism when the value of the coupling constant satisfies $10^{-6}\leq \tilde{g}\leq 3\times 10^{-5}$, { what improves the result $10^{-6}\leq \tilde{g}\ll 1$ obtained in \cite{fkl}}. For example, taking $\tilde{g}=6\times 10^{-6}$, we obtain
\begin{eqnarray}\label{reheating-decay}
    T_{reh}\cong 5\times 10^{-13}\left(
\frac{M_{pl} }{\bar\Gamma}
    \right)^{1/4}\ln^{3/4}\left( \frac{M_{pl}}{\bar\Gamma}\right)M_{pl}, \qquad \mbox{with} \qquad 10^{-6}
    \leq  \frac{\bar\Gamma}{M_{pl}} <1/3,\end{eqnarray}
what leads to the following 
 maximum and minimum reheating temperatures
\begin{eqnarray}\label{max-min}
T_{reh}^{max}\cong  10^{-10} M_{pl} \qquad \mbox{and}\qquad
T_{reh}^{min}\cong 7\times 10^{-13} M_{pl}.
\end{eqnarray}

\

In summary, choosing $\tilde{g}\cong 6\times 10^{-6}$ guarantees that the vacuum polarization effects do not disturb the evolution of the inflaton field during the last stages of inflation. Additionally, the particles become non-relativistic and have masses less than $M_{pl}$ soon after the start of kination, and their decay occurs during this phase. In this situation, a viable reheating temperature below $5\times 10^{10} M_{pl}$ is obtained, ensuring the success of BBN.

\section{Number of e-folds}

Let $N$ be the number of e-folds from horizon crossing to the end of inflation. Then, we have
\begin{equation}
a_* = e^{-N}a_{{END}},
\end{equation}
where $"END"$ denotes the end of inflation, and once again, the "star" means that the quantities are evaluated at horizon crossing. Since the pivot scale $k_*$ is defined as $k_* = a_*H_*$ (at horizon crossing), we have
\begin{eqnarray}
    \frac{k_*}{a_0H_0}=e^{-N}\frac{H_*}{H_0}\frac{a_{END}}{a_{kin}}
    \frac{a_{kin}}{a_{end}} \frac{a_{end}}{a_{m}}    \frac{a_{m}}{a_{0}}
    =e^{-N}\frac{H_*}{H_0}\frac{a_{END}}{a_{kin}}
    \frac{\rho_{end}^{-1/12}\rho_m^{1/4}}{\rho_{kin}^{1/6}}\frac{a_{m}}{a_{0}}    ,     \end{eqnarray}
where  the sub-index $"m"$ denotes the matter-radiation equality,  "end" the end of kination,  $"0"$
the present time, and we have used the relations
\begin{eqnarray}
    \rho_{end}=\rho_{kin}\left(\frac{a_{kin}}{a_{end}} \right)^6,  \qquad 
    \rho_{m}=\rho_{end}\left(\frac{a_{end}}{a_{m}} \right)^4.\end{eqnarray}

\

In dealing with Instant Preheating, we have already shown that the decay of non-relativistic particles must occur prior to the end of kination. Therefore, we will have:
\begin{eqnarray}
    \rho_{end}=\rho_{reh}= \frac{g_{reh}\pi^2}{30}T_{reh}^4.
\end{eqnarray}

Next, as a physical scale, we use $k_{\text{phys},0}\equiv k_*/a_0=2\times 10^{-2}\text{ Mpc}^{-1}$ \cite{planck18}, and for the current Hubble scale, $H_0\cong 2\times 10^{-4}\text{ Mpc}^{-1}\cong 6\times 10^{-61}M_{pl}$. In addition, since the evolution is adiabatic after the matter-radiation equality, i.e., entropy is conserved, we have
$
a_mT_m=a_0T_0,
$
as well as the relation
$
\rho_{m}= \frac{g_{m}\pi^2}{30}T_{m}^4,
$
where $g_m=3.36$ is the effective number of degrees of freedom at the matter-radiation equality. Hence,
\begin{eqnarray}
    N=-4.6+
    \ln\left(\frac{H_*}{H_0} \right)+\ln\left(\frac{a_{END}}{a_{kin}} \right)
    +\frac{1}{6}\ln\left(\frac{\rho_{reh}}{\rho_{kin}} \right)  
    +\frac{1}{4}\ln\left(\frac{g_{m}}{g_{reh}} \right)  
    +\ln\left(\frac{T_0}{T_{reh}} \right).    \end{eqnarray}

Now, considering the formula for the power spectrum of scalar perturbations (\ref{power}), we can infer that
$H_* \approx 4 \times 10^{-4} \sqrt{\epsilon_*} M_{pl}$.
By using the present values of the Hubble rate and temperature   $T_0 \approx 2.73 \ \text{K} \approx 2 \times 10^{-13} \ \text{GeV} \approx 8 \times 10^{-32} M_{pl}$, we can calculate the number of e-folds as a function of the reheating temperature and $\epsilon_*$.
\begin{eqnarray}\label{Nxx}
    N(T_{reh}, \epsilon_*)
    \cong 54.47 +\frac{1}{2}    \ln \epsilon_*  
    +\frac{1}{3}\ln \left( \frac{M_{pl}^2}{T_{reh}H_{END}}\right),   \end{eqnarray}
where we have neglected the model-dependent term $\ln\left(\frac{a_{END}}{a_{kin}} \right)$ since it is close to zero, and we have assumed that there is no significant drop in energy during the phase transition from the end of inflation to the beginning of kination.

It is important to note that for a given potential $V$, one can calculate
\begin{eqnarray}
    \epsilon_*=\frac{M_{pl}^2}{2}\left( \frac{V'_*}{V_*}\right)^2 \qquad
    \mbox{and} \qquad H_{END}^2=\frac{V_{END}}{2M_{pl}^2}.
\end{eqnarray}

\

On the other hand,  the number of efolds can also be calculated from the formula
\begin{eqnarray}
    N\cong \frac{1}{M_{pl}^2}\int_{\varphi_*}^{\varphi_{END}} 
    \left|\frac{V(\varphi)}{V'(\varphi)}\right|d\varphi    =\frac{1}{M_{pl}}\int_{\varphi_*}^{\varphi_{END}}\frac{1}{\sqrt{2\epsilon}}d\varphi. 
\end{eqnarray}

As we will see, is a function of $\epsilon_*$, which is also related with 
 the spectral index $n_s$.  By equating both expressions, we obtain a relationship between the reheating temperature and the spectral index, given by:
$N(T_{reh}, \epsilon_*(n_s))=N(n_s)$.
We will explore this relationship in the next section.

\section{Reheating constraints}

In this section, we will use the results obtained in the previous section to analyze the feasibility of three important Quintessential Inflation models. Specifically, we will study the relationship between the reheating temperature and the spectral index for each of these models. By analyzing these relationships, we can determine if these models are consistent with observational data and if they are viable candidates for explaining the evolution of the  Universe.

\subsection{The Peebles-Vilenkin model}

The first Quintessential Inflation scenario was proposed by Peebles and Vilenkin in their seminal paper \cite{pv} at the end of the 20th century, shortly after the discovery of cosmic acceleration. The corresponding potential is given by
\begin{eqnarray}\label{pv}
V(\varphi)=\left\{\begin{array}{ccc}
\lambda (\varphi^4+M^4)& \mbox{for}& \varphi\leq 0\\
\lambda \frac{M^8}{\varphi^4+M^4} &\mbox{for}& \varphi\geq 0.
\end{array}\right.
\end{eqnarray}

Here, $\lambda\sim 10^{-14}$ is a dimensionless parameter and $M$ is a very small mass compared to the Planck mass $M_{pl}$.
It is important to note that the quartic potential is responsible for inflation, while the inverse power law leads to dark energy (in this case, quintessence) at later times.

\

Since for this model 
$\epsilon=\frac{8M_{pl}^2}{\varphi^2}$,  we have $\varphi_{END}=-2\sqrt{2}M_{pl}$, and 
taking into account that for a quartic potential  $3\epsilon_*=1-n_s$, we get  $\varphi_*=-\frac{2\sqrt{6}}{\sqrt{1-n_s}}M_{pl}$. So,   the number  of efolds
will be
\begin{eqnarray}\label{N-pv}
    N=\frac{1}{4M_{pl}^2}(\varphi_*^2-\varphi_{END}^2)=\frac{6}{1-n_s}-2.
    \end{eqnarray}

On the other hand,
 using  that inflation ends when
$\epsilon_{END}=1$, i.e., when  $w_{eff}=-1/3$,
one has  $\dot{\varphi}_{END}^2=V(\varphi_{END})$,
and thus, 
\begin{eqnarray}
    \rho_{END}=\frac{3V(\varphi_{END})}{2}=96\lambda M_{pl}^4 \Longrightarrow H_{END}=4\sqrt{2\lambda}M_{pl}.
\end{eqnarray}

 Then, from Eqs. (\ref{N-pv}) and (\ref{Nxx}) we get
\begin{eqnarray}
    \frac{6}{1-n_s}-\frac{1}{2}\ln\left(\frac{1-n_s}{3}\right)\cong 56.47+
    \frac{1}{3}\ln\left(\frac{M_{pl}}{4\sqrt{2\lambda}T_{reh} } \right),
\end{eqnarray}
which for $\lambda\cong 10^{-14}$,  leads to 
\begin{eqnarray}
    T_{reh}\cong (1-n_s)^{3/2}\exp\left(182-\frac{18}{1-n_s}\right) M_{pl}.
\end{eqnarray}

Finally, expressing the reheating temperature $T_{reh}$ as a function of the spectral index $n_s$, we can use observational data to constrain the parameter space. According to Planck 2018 data, the spectral index is measured to be $n_s=0.9649\pm 0.0042$ \cite{planck18}. At the 2$\sigma$ Confidence Level, the minimum value of $n_s$ that leads to the maximum reheating temperature is $n_s=0.9565$. However, for this value of $n_s$, the reheating temperature is found to be abnormally small:
\begin{eqnarray}
T_{reh}\sim 10^{-2}e^{-234}M_{pl}.
\end{eqnarray}
This demonstrates that the Peebles-Vilenkin model is not viable, as it predicts an unreasonably low reheating temperature for the observed spectral index.

\

Equivalently, the non-feasibility of the Peebles-Vilenkin model can also be seen by calculating the number of e-folds using Eq. (\ref{N-pv}). At the 2$\sigma$ confidence level, this leads to the bound $136\leq N\leq 223$, which is in contradiction with the number of e-folds calculated from Eq. (\ref{Nxx}). Using Eq. (\ref{Nxx}) and requiring a reheating temperature above $1$ MeV and below $10^9$ GeV, the number of e-folds is constrained to satisfy $63\leq N\leq 74$. Therefore, the Peebles-Vilenkin model is not viable because it predicts a number of e-folds that is outside of the observational constraints.

\

{

A final remark is in order: The latest observational data constrain the tensor-to-scalar ratio of scalar perturbations, $r$, to be less than $0.1$. For the Peebles-Vilenkin model, one has $r=\frac{16}{3}(1-n_s)$, and taking into account that $n_s=0.9649\pm 0.0042$ at $2\sigma$ C.L., one has the constraint $0.1424\leq r\leq 0.232$, which is incompatible with the observational bound $r\leq 0.1$. This provides another way to show that this model is not viable. The difference with our methodology is that we do not need a precise bound on the tensor-to-scalar ratio to disregard this model.

}

\subsection{Exponential $\alpha$-attractor}

We consider   
a Quintessential Inflation  $\alpha$-attractor model, whose potential is given by \cite{haro21}
\begin{eqnarray}\label{alpha}
V(\varphi)=\lambda M_{pl}^4e^{-n\tanh\left(\frac{\varphi}{\sqrt{6\alpha}M_{pl}} \right)},
\end{eqnarray}
where $\lambda$, $\alpha$ and $n$ are some dimensionless parameters.
The value of the slow roll parameter $\epsilon$ is
\begin{eqnarray}
    \epsilon=
    \frac{n^2}
    {12\alpha}\frac{1}{\cosh^4\left(\frac{\varphi}{\sqrt{6\alpha}M_{pl}} \right)},
\end{eqnarray}
and the other slow-roll parameter  is given by
\begin{eqnarray}
    \eta
    =
    \frac{n}
    {3\alpha}\left[\frac{\tanh\left(\frac{\varphi}{\sqrt{6\alpha}M_{pl}} \right)}{\cosh^2\left(\frac{\varphi}{\sqrt{6\alpha}M_{pl}} \right)} +
    \frac{n/2}{\cosh^4\left(\frac{\varphi}{\sqrt{6\alpha}M_{pl}} \right)}    
     \right].
\end{eqnarray}

Both slow-roll parameters must be evaluated at the horizon crossing, which occurs for large values of 
$\cosh\left(\frac{\varphi}{\sqrt{6\alpha}M_{pl}} \right)$, obtaining
\begin{eqnarray}
    \epsilon_*=
    \frac{n^2}
    {12\alpha}\frac{1}{\cosh^4\left(\frac{\varphi_*}{\sqrt{6\alpha}M_{pl}}  \right)} \qquad \mbox{and} \qquad
\eta_*\cong 
    -\frac{n}
    {3\alpha}\frac{ 1}{\cosh^2\left(\frac{\varphi_*}{\sqrt{6\alpha}M_{pl}} \right)}, 
     \end{eqnarray}
with $\varphi_*<0$. Therefore, the number of efolds is
\begin{eqnarray}
    N\cong 
    \frac{6\alpha}{n}
    \cosh^2\left(\frac{\varphi_*}{\sqrt{6\alpha}M_{pl}}\right)
    \cong 
    \sqrt{\frac{3\alpha}{4\epsilon_*}},
\end{eqnarray}
which is related with the spectral index of the scalar perturbations via the relation
\begin{eqnarray}
    n_s-1\cong -6\epsilon_*+2\eta_*\cong 2\eta_*
=- \frac{4\sqrt{\epsilon_*}}{\sqrt{3\alpha}}\cong -\frac{2}{N},
\end{eqnarray}
obtaining
\begin{eqnarray}\label{spectral}
    N(n_s)\cong \frac{2}{1-n_s} \qquad \mbox{and} \qquad \epsilon_*(n_s)\cong \frac{3\alpha}{16}(1-n_s)^2.
\end{eqnarray}

\

From Eq. (\ref{spectral}), we can also calculate the relationship between the parameters of the model. Specifically, since $V_*\cong \lambda M_{pl}^4 e^n$, we have
$H_*^2\cong \frac{\lambda M_{pl}^2}{3}e^n$. Thus, using the formula for the power spectrum of scalar perturbations, we obtain the constraint:
\begin{eqnarray}\label{alpha-constraint}
    \frac{\lambda}{\alpha \pi^2(1-n_s)^2}e^n\cong 9\times 10^{-9}.
\end{eqnarray}

\

We also need to calculate $H_{END}$, which can be done by noting that $\epsilon_{END}=1$ and using 
$
\mathop{\mathrm{arccosh}}(x)=\ln(x-\sqrt{x^2-1}),
$
to obtain
\begin{eqnarray}
    \varphi_{END}=\sqrt{6\alpha}\ln\left( \frac{\sqrt{n}}{(12\alpha)^{1/4}}-\sqrt{\frac{n}{\sqrt{12\alpha}}-1}
    \right) M_{pl}^4.
\end{eqnarray}

Inserting it in (\ref{alpha}) and using the constraint (\ref{alpha-constraint}), one has
\begin{eqnarray}
V(\varphi_{END})=\lambda M_{pl}^4e^{n\sqrt{1-\frac{\sqrt{12\alpha}}{n}}}\cong \lambda M_{pl}^4e^{n\left(1-\frac{\sqrt{3\alpha}}{n}\right)}\cong 9\pi^2\alpha (1-n_s)^2e^{-\sqrt{3\alpha}}10^{-9} M_{pl},
\end{eqnarray}
and thus, 
\begin{eqnarray}\label{H}
\rho_{END}=\frac{3V(\varphi_{END})}{2} \Longrightarrow H_{END}\cong 3\sqrt{\frac{\alpha}{2}}(1-n_s)e^{-\sqrt{3\alpha}/2} 10^{-4} M_{pl}.
    \end{eqnarray}

\

Therefore, by equating both expressions for the number of e-folds, we obtain an expression for the reheating temperature as a function of the spectral index
\begin{eqnarray}\label{reheating-alpha}
    T_{reh}\cong \alpha (1-n_s)^2 \exp\left(169
    +\frac{\sqrt{3\alpha}}{2}    -\frac{6}{1-n_s}\right) M_{pl}.
\end{eqnarray}

To be more precise,  we will choose $\alpha=10^{-2}$, obtaining
\begin{eqnarray} \label{alpha-0.01}    T_{reh}\cong  (1-n_s)^2 \exp\left(169+\frac{\sqrt{3}}{20}
      -\frac{6}{1-n_s}\right) 10^{-2} M_{pl}.
\end{eqnarray}

We can check that the allowed values of the spectral index, which lead to a reheating temperature compatible with the one obtained in Eq. (\ref{max-min}) using Instant Preheating, are in the range of $(0.9667, 0.9677)$. Thus, the value of the spectral index is approximately $n_s\cong 0.967$. Additionally, the ratio of tensor to scalar perturbations is given by $r=16\epsilon_*=3\alpha(1-n_s)^2$, and we can conclude that its value is $r\cong 3\times 10^{-5}$.

\subsubsection{Comparison with other works}

{

In \cite{dimopoulos}, the authors study an exponential $\alpha$-attractor with a cosmological constant given by
\begin{eqnarray}\label{alpha1}
V(\varphi)=\lambda M_{pl}^4 \left(e^{-n\tanh\left(\frac{\varphi}{\sqrt{6\alpha}M_{pl}} \right)}-e^{-n}\right),
\end{eqnarray}
which at early times coincides with our potential (\ref{alpha}), but at late times, it will become
\begin{eqnarray}\label{alpha2}
V(\varphi)=2n\lambda e^{-n}M_{pl}^4e^{-\sqrt{\frac{2}{3\alpha}}\varphi/M_{pl}}.
\end{eqnarray}

\

This constrains the value of the parameter $\alpha$ to match with the current Planck data of the effective equation of state (EoS) parameter for dark energy. As shown in the Appendix of \cite{dimopoulos}, the parameter $\alpha$ must satisfy $\alpha \geq 3/2$ (although it has been shown in \cite{linde} that the correct bound to match the observational data is $0.5\leq \alpha \leq 3.3$). Fortunately, this is not a problem for our model, as demonstrated in \cite{haro21} and in \cite{linde} where the authors found that $\alpha$ only has to satisfy the upper bound $\alpha<3.5$. For the value chossen in this work $\alpha = 10^{-2}$, the present value of the effective EoS parameter is approximately $-0.68$, which is compatible with the Planck data. Additionally, from the observational data $\Omega_{\varphi,0}=\frac{V(\varphi_0)}{3H_0^2M_{pl}^2}\cong 0.7$ (where the subscript "0" denotes present time), \cite{dimopoulos} obtains a relationship between the parameters $\alpha$, $n$, and $\tilde{g}$. This is because the present value of the scalar field depends on the reheating temperature, which in turn depends on $\tilde{g}$. Looking at equation (\ref{alpha2}), we can see that $V(\varphi_0)$ depends on all three parameters.

However, this is not the case for our potential (\ref{alpha}). At the present time, $V(\varphi_0)\sim \lambda M_{pl}^4 e^{-n}$. Thus, $\Omega_{\varphi,0}\cong 0.7$ leads to $\lambda e^{-n}\cong 10^{120}$. Combining this with (\ref{alpha-constraint}), we obtain the relationship
\begin{eqnarray}
e^{2n}\cong 9\alpha\pi^2 (1-n_s)^2\times 10^{111},
\end{eqnarray}
which is independent of $\tilde{g}$. In fact, since we have obtained $n_s\cong 0.967$ for $\alpha=10^{-2}$, we find $n\cong 124$ and $\lambda\cong 10^{-66}$.

\

On the other hand, in contrast to our realistic assumption that the particles produced during kination decay when they are non-relativistic, in \cite{dimopoulos}, it is assumed that the decay of these particles occurs immediately after their creation. We find it difficult to justify this assumption because, at the onset of kination, the effective mass of the produced particles vanishes. Additionally, it is assumed that the produced particles decay into light fermions with a decay rate given by $\Gamma=\frac{h^2 m_{eff}}{8\pi}$ (as argued in \cite{fkl}, where the authors suggest that the decay should occur when the particles are non-relativistic). However, the decay that occurs when $H\sim \Gamma$ cannot happen at the beginning of kination because, at that time, $\Gamma\cong 0$ and $H_{kin}\sim 10^{-7}M_{pl}$. Nonetheless, it is possible that another kind of decay may occur.

\

Assuming that the decay occurs immediately after the beginning of kination, the reheating temperature can be calculated using the simple formula:

\begin{eqnarray}\label{reheating-immediate}
T_{reh}=\left( \frac{270}{\pi^{11}g_{reh}}\right)^{1/4}\tilde{g}^{3/2}\sqrt{H_{kin}M_{pl}}\cong 2\times 10^{-5}\tilde{g}^{3/2}
M_{pl},
\end{eqnarray}

where $\langle\hat{\rho}_{kin}\rangle$ is given by:

\begin{eqnarray}
\langle\hat{\rho}_{kin}\rangle=\frac{1}{2\pi^2a_{kin}^4}\int_0^{\infty}k^3|\beta_k|^2dk=
\frac{\tilde{g}^2\dot{\varphi}^2_{kin}}{8\pi^3},
\end{eqnarray}

with $\dot{\varphi}^2_{kin}=6H_{kin}^2M_{pl}^2$, and we have taken $H_{kin}\cong 10^{-7} M_{pl}$. It is worth noting that the formula (\ref{reheating-immediate}) depends solely on $\tilde{g}$, in contrast to the formula (\ref{reheating-decay}), which depends on both $\tilde{g}$ and the decay rate $\Gamma$.
Additionally,  the bounds arising from the gravitino constraint ($T_{reh} \leq 10^9$ GeV) and the lower bound $T_{reh} \geq 1$ MeV,  constraint in different ways the parameter $\tilde{g}$. Effectively, when the decay is at the onset of kination one has $10^{-4}\leq \tilde g \leq 10^{-2}$, but if it occurs when the particles are non-relatistic, as we have already shown,  this parameter has to satisfy $10^{-6}\leq \tilde{g}\leq 3\times 10^{-5}$.

\

Therefore, in accordance with \cite{dimopoulos}, one must select, in the corresponding allowed range,  values for the parameters $\alpha$ and $n$ that result in a value of $\tilde{g}$ (since these three parameters are related for the model (\ref{alpha1})) that is consistent with the constraints arising from the gravitino constraint  and the lower bound of the reheating temperature. Once this value of $\tilde{g}$ is determined, the reheating temperature can be calculated using formula (\ref{reheating-immediate}), and the spectral index can be calculated using the relationship between the reheating temperature and the spectral index. In summary, for the set of allowed parameters $\alpha$ and $n$, it is possible to calculate the reheating temperature and the spectral index by choosing appropriate values for the coupling constant $\tilde{g}$.

\

This is quite different from our method. In our approach, we use instant preheating with the realistic assumption that the created particles can only decay when they become non-relativistic, which results in a range of reheating temperatures for the potential (\ref{alpha}). Therefore, after fixing the model and setting $\alpha=10^{-2}$ to satisfy the gravitino constraint and ensure that the vacuum fluctuations do not disturb the evolution of the scalar field during the last stages of inflation, the value of $\tilde{g}$ has to belong to a very narrow range. Once we have fixed the value of $\tilde{g}$, we use formula (\ref{reheating-decay}) to obtain the range of viable values of the reheating temperature. Finally, by using the relationship between the reheating temperature and the spectral index, we can determine the range of viable values of the spectral index for the fixed model.

\

Another paper that deals with $\alpha$-attractors is \cite{linde}, where the authors study several potentials and use observational data to constrain the parameters of each model. The work does not deal with any preferred reheating mechanism, but it compares instant preheating with reheating via gravitational particle production of light particles, showing that instant preheating is more efficient because the gravitational production of light particles leads to a low reheating temperature of the order of $10^5$ GeV (see, for instance, \cite{pv}). Furthermore, it is pointed out that due to the kination phase, the number of last e-folds is greater in Quintessential Inflation than in standard inflation. As a consequence, the value of the spectral index is greater in Quintessential Inflation than in standard models. Thus, future improvements in the accuracy of the measurement of the spectral index may distinguish between conventional inflationary models with a cosmological constant and Quintessential Inflation scenarios.

\

Finally, in \cite{linde}, the parameters of the models (\ref{alpha}) and (\ref{alpha1}) are compared with observational data.
Only taking into account that for lower reheating temperatures, the inflaton field freezes later during radiation than for higher reheating temperatures. This means that when reheating is via gravitational production of light particles, the inflaton field freezes later than in the case when the reheating mechanism is Instant Preheating.
 Since the models have completely different tails, this leads to different equations of state (EoS) parameters at late times, depending on the freeze value of the inflaton field, and thus on the value of the reheating temperature. This constrains the values of $\alpha$. Specifically, for the model described by equation (\ref{alpha}), the value of $\alpha$ is bounded by $\alpha<3.5$, and therefore our choice of $\alpha=10^{-2}$ is entirely acceptable. In contrast, for the model described by equation (\ref{alpha1}), the observational data only allows values within the range of $0.5\leq \alpha\leq 3.3$.}

\subsection{The double exponential  model}

Next, we consider a combination of two exponential potentials to depict inflation and quintessence, respectively:
\begin{eqnarray}\label{viable}
V(\varphi)= V_0e^{-\bar{\gamma}{\varphi}^n/{M_{pl}^n}}+ M^4e^{-\gamma \varphi/M_{pl}},
\end{eqnarray}
where we must choose $0<\gamma<\sqrt{2}$ to model the current cosmic acceleration. This is because, at late times, the effective equation of state parameter is $w_{eff}=\frac{\gamma^2}{3}-1<-1/3$.

\

The first part of the potential { is a phenomenological term responsible for inflation}, which   has been  studied in detail  in \cite{Geng,hossain}, where it is obtained that 
\begin{eqnarray}\label{epsilon-double}
\epsilon
=\frac{\bar\gamma^2n^2}{2}\left( \frac{\varphi}{M_{pl}} \right)^{2n-2}.
\end{eqnarray}

So, 
at the end of inflation  one has
$\varphi_{END}=\left( \frac{2}{n^2\bar\gamma^2}\right)^{\frac{1}{2n-2}} M_{pl}$, 
and thus,  
{
 \begin{eqnarray}\label{HEND}
 \rho_{\varphi, END}=\frac{3}{2} V(\varphi_{END})\cong 9\pi^2 e^{-\bar\gamma\left( \frac{2}{n^2\bar\gamma^2}\right)^{\frac{n}{2n-2}}}
 \times 10^{-11} M_{pl}^4\Longrightarrow H_{ END}\cong \sqrt{\frac{3}{10}}
  e^{-\frac{\bar\gamma}{2} \left( \frac{2}{n^2\bar\gamma^2}\right)^{\frac{n}{2n-2}}}
 \times 10^{-5} M_{pl}, \end{eqnarray}}
which will constrain the values of the parameter $\bar\gamma$ significantly, because in all viable inflationary models, the value of the Hubble rate at the end of inflation is of the order of $10^{-6} M_{pl}$. In fact, when (\ref{HEND}) is of the order of $10^{-6} M_{pl}$, we get:
 {
\begin{eqnarray}\label{bargamma}
    \bar{\gamma}n=\sqrt{2}\left(\frac{\sqrt{2}}{3n}\right)^{n-1}.
\end{eqnarray}

Next,  we calculate the other slow-roll parameter
\begin{eqnarray}
    \eta=-\frac{2(n-1)}{3n} \epsilon^{\frac{n-2}{2n-2}}+2\epsilon,
\end{eqnarray}
leading to the following spectral index
\begin{eqnarray}\label{double-spectral}
    1-n_s=2\epsilon_*+\frac{4(n-1)}{3n}\epsilon_*^{\frac{n-2}{2n-2}}.
    \end{eqnarray}

On the other hand, for $n>2$,  the number of efolds is given by
\begin{eqnarray}\label{N-efolds-double}
{ N}=
\frac{1}{n\bar\gamma(n-2)}
\left[\left( \frac{\varphi_*}{M_{pl}} \right)^{2-n}-
\left( \frac{\varphi_{END}}{M_{pl}} \right)^{2-n}
\right]
=\frac{3n}{2(n-2)}
\left[ \epsilon_*^{\frac{2-n}{2n-2}}-1
\right]
\cong\frac{3n}{2(n-2)} \epsilon_*^{\frac{2-n}{2n-2}}
\cong\frac{2(n-1)}{n-2}\frac{1}{1-n_s},
\end{eqnarray}
where we have used the approximation
$ 1-n_s \cong \frac{4(n-1)}{3n}\epsilon_*^{\frac{n-2}{2n-2}}$.
}

\

Therefore, from the equations (\ref{Nxx}), (\ref{double-spectral}), and (\ref{N-efolds-double}), we obtain the reheating temperature as a function of the spectral index:

\begin{eqnarray}
T_{reh}=\left( \frac{3n}{4n-4}\right)^{3/2}(1-n_s)^{\frac{3n-3}{n-2}}
\exp\left( 177.21-\frac{6n-6}{(n-2)(1-n_s)} \right) M_{pl}.
\end{eqnarray}

Note that the maximum reheating temperature is obtained from the minimum observable value of the spectral index, which at $2\sigma$ C.L. is $n_s=0.9565$. Consequently,

\begin{enumerate}
\item For $n=3$, the maximum reheating temperature is of the order of $10^{-51} M_{pl}$.
\item For $n=4$, the maximum reheating temperature is of the order of $9\times 10^{-20} M_{pl}$.
\item For $n=5$, the maximum reheating temperature is of the order of $4\times 10^{-9} M_{pl}$.
\item For $n\gg 1$, the maximum reheating temperature is of the order of $6\times 10^{12} M_{pl}$.
\end{enumerate}

In the same way, the maximum observable value of the spectral index, which at $2\sigma$ C.L. is $n_s=0.9733$, leads to the minimum reheating temperature. In the double exponential model, it is below $10^9$ GeV when $n>2$. Thus, since a viable model has to satisfy that the maximum temperature is above $1$ MeV and the minimum one below $10^9$ GeV, we can conclude that the viable double exponential models, those that have a range of values of the spectral index leading to a reheating temperature between $1$ MeV and $10^9$ GeV, are the ones satisfying $n>3$.

\

To end the Section, 
when  $n\gg 1$, the reheating temperature is approximately
\begin{eqnarray}
    T_{reh}\cong \left( \frac{3}{4}\right)^{3/2}(1-n_s)^3
    \exp\left( 177.21-\frac{6}{(1-n_s)} \right) M_{pl},
\end{eqnarray}
and the viable values of the spectral index 
compatibles with the reheating temperature via Instant Preheating (\ref{max-min}),  are 
 in the range  $0.9683\leq n_s\leq 0.9688$, i.e., $n_s\cong 0.9685$
 and $r=9(1-n_s)^2\cong9\times 10^{-3}$.

\section{Reheating via gravitational particle production}

When reheating is produced via gravitational particle production of heavy particles whose decay is before the end of kination, the reheating temperature is given by \cite{haro23}:
\begin{eqnarray}\label{reheating3}
 T_{{reh}}=  
 \left(\frac{10}{3\pi^2g_{reh}} \right)^{1/4}
 \left(\frac{\langle\hat\rho_{kin}\rangle^3}{H_{kin}^3
 \Gamma M_{pl}^8}\right)^{1/4}M_{pl},
 \end{eqnarray}
where the decay rate $\Gamma$ has to be within the following range:
\begin{eqnarray}\label{constraint-gamma}
    \frac{\langle \hat\rho_{kin}\rangle}{3H_{kin}M_{pl}^2}\leq \Gamma\leq H_{kin}.
\end{eqnarray}

The maximum reheating temperature is reached { at the end of kination, i.e.,} when 
$\frac{\langle \hat\rho_{kin}\rangle}{3H_{kin}M_{pl}^2}= \Gamma$. Therefore,  
\begin{eqnarray}\label{Tmax}
    T_{reh}^{max}= \left( \frac{10}{\pi^2 g_{reh}} \right)^{1/4}\sqrt{  \frac{\langle \hat\rho_{kin}\rangle}{H_{kin}M_{pl}^3}}M_{pl}.\end{eqnarray}

\

According to \cite{haro23}, it has been demonstrated that the energy density of conformally coupled particles created at the onset of kination can be approximated by the analytical formula
\begin{eqnarray}\label{rho1}
   \langle \hat\rho_{kin}\rangle
   \cong
   \frac{1}{4\pi^3}
   e^{-\frac{\pi m_{\chi}}{2\sqrt{2}H_{END}}}
   \sqrt{\frac{m_{\chi}}{\sqrt{2}H_{END}}} H_{END}^2m_{\chi}^2, \end{eqnarray}
where $m_{\chi}$ is the mass of the produced particles. Inserting this expression in (\ref{Tmax}) we get
\begin{eqnarray}\label{max-temp}
    T_{{reh}}^{{max}}(m_{\chi})\cong
    2\times 10^{-2}e^{-\frac{\pi m_{\chi}}{4\sqrt{2}H_{END}}}    
    \left( \frac{m_{\chi}H_{END}}{M_{pl}^2} \right)^{1/4}
    m_{\chi}.\end{eqnarray}

By applying the previous result to the exponential $\alpha$-attractor model with $\alpha=10^{-2}$, we can insert equation (\ref{max-temp}) into equation (\ref{alpha-0.01}) to establish a relationship between the spectral index and the mass of the produced particles
\begin{eqnarray}\label{alpha-alpha}
    \frac{2.6057}{1-n_s}-1.75\log(1-n_s)=79.301+\frac{1.2398}{1-n_s}X-1.25\log X,
\end{eqnarray}
where we have introduced the notation $X\equiv 10^4 \frac{m_{\chi}}{M_{pl}}$.

\

It is important to note that the equation (\ref{alpha-alpha}) has a solution for a minimum value of $n_s$, which is obtained at the minimum of the function $f(X)=\frac{1.2398}{1-n_s}X-1.25\log X$. By inserting $X_{min}=\frac{1.25(1-n_s)}{1.2398}$ into (\ref{alpha-alpha}), we obtain:
\begin{eqnarray}\label{xxxx}
    \frac{2.6057}{1-n_s}-0.5\log(1-n_s)=80.5465.\end{eqnarray}

The only solution to this equation is $\bar{n}s\cong0.9673$ because the function
\begin{eqnarray}
\frac{2.6057}{1-n_s}-0.5\log(1-n_s),
\end{eqnarray}
is increasing. This implies that Eq. (\ref{xxxx}) has only one solution. For this minimum value of the spectral index, the equation (\ref{alpha-alpha}) also has a unique solution: $m{\chi}\cong 10^{-6}M_{pl}$, which leads to a maximum reheating temperature of around $10^7$ GeV

\

For values of the spectral index in the range $(0.9673, 0.9709)$ (where $n_s=0.9709$ is the maximum value leading to a reheating temperature above $1$ MeV), equation (\ref{alpha-alpha}) has two solutions. For example, when $n_s=0.9709$, there are two compatible masses: $m_{\chi}\cong 3\times 10^{-5} M_{pl}$ and $m_{\chi}\cong 5\times 10^{-15}M_{pl}$, with a reheating temperature of $1$ MeV.

\

In other words, assuming that the decay occurs at the end of kination (which leads to the maximum reheating temperature), the allowed values of the spectral index are in the range $(0.9673, 0.9709)$, and for each of these values, there are two compatible masses. Alternatively, for masses satisfying the inequality $5\times 10^{-15}\leq m_{\chi}/M_{pl}\leq 3\times 10^{-5}$, there is a value of the spectral index in the range $0.9673<n_s<0.9709$, which leads to a viable maximum reheating temperature.

\section{Conclusions}

Throughout this work, we have emphasized the close relationship between the reheating temperature and the spectral index of scalar perturbations. We have explored this connection in the context of non-oscillating cosmologies, where the reheating mechanism is the well-known Instant Preheating. Our analysis has shown that the viable range of reheating temperatures falls between $10^5$ GeV and $10^8$ GeV, resulting in a narrow range of viable values for the spectral index. Specifically, for an exponential $\alpha$-attractor with $\alpha=10^{-2}$, we find that the spectral index is close to $n_s\cong 0.9670$, while for a double exponential model, we obtain $n_s\cong 0.9685$.

\

{

 We have also compared the methodology used in this work with that of \cite{dimopoulos}. We pointed out that the main difference between the two approaches is that, in our work, we make the realistic assumption (following the spirit of \cite{fkl}) that the created particles must decay when their effective mass is great enough to be considered non-relativistic. This assumption does not hold in \cite{dimopoulos}, where the authors assume that the particles decay immediately after their creation, resulting in a vanishing effective mass. This leads to differences in the results obtained and in the parameter constraints.

}

\

Finally, 
we have also considered an alternative to Instant Preheating: reheating via gravitational particle production. In this case, we have established a connection between the spectral index and the masses of the produced particles. Our analysis shows that the production of heavy particles with masses less than $10^{-5} M_{pl}$ can lead to viable reheating temperatures, and values of the spectral index that fall within the observational domain provided by the Planck team, at $2\sigma$ C.L.

\

\section*{Acknowledgments} 
This work 
is supported by the Spanish grant 
PID2021-123903NB-I00
funded by MCIN/AEI/10.13039/501100011033 and by ``ERDF A way of making Europe''.

\

\section*{Appendix: The diagonalization method}

Expanding a quantum field conformally coupled with gravity in terms of the creation and annihilation operators
\begin{eqnarray}\label{expansion}
    \hat{\phi}(\eta, {\bf x})=\frac{1}{(2\pi)^{3/2}a(\eta)}
    \int_{\mathbb R^3} (\hat{a}_{\bf k}
    \chi_k(\eta)e^{i{\bf k}{\bf x}}
    +\hat{a}_{\bf k}^{\dagger}\bar\chi_k(\eta)e^{-i{\bf k}{\bf x}}    
    ) d^3{\bf k},\end{eqnarray}
where 
 $\chi_k$ and its conjugate $\bar\chi_k$ are the mode solution of the Klein-Gordon equation 
\begin{eqnarray}\label{KG}
    \chi_k''+\omega_k^2(\eta)\chi_k=0,\end{eqnarray}
 with initial conditions, at some early time $\eta_i$,  
 \begin{eqnarray}
 \chi_k(\eta_i)=\frac{1}{\sqrt{2\omega_k(\eta_i)}} \qquad  \mbox{and}
 \qquad
\chi_k'(\eta_i)=-i\sqrt{\frac{{\omega_k(\eta_i)}}{2}}.
\end{eqnarray}

\

The renormalized vacuum energy density is given by
\begin{eqnarray}\label{vacuum-energy1}\langle {\hat\rho}(\eta)\rangle=\frac{1}{4\pi^2a^4(\eta)}\int_0^{\infty} k^2dk \left(   |\chi_{ k}'(\eta)|^2+ \omega^2_k(\eta) |\chi_{ k}(\eta)|^2-  \omega_k(\eta)\right),\end{eqnarray}
where we have subtracted the zero point oscillations of the vacuum.

\

To express the vacuum energy density in a simple form, we can use the diagonalization method, which involves expanding the modes as follows \cite{Zeldovich} (see also Section $9.2$ of \cite{gmmbook}):
\begin{eqnarray}\label{zs}
\chi_{k}(\eta)= \alpha_k(\eta)\phi_{k,+}(\eta)+
\beta_k(\eta)\phi_{k,-}(\eta),
\end{eqnarray}
where $\alpha_k(\eta)$ and $\beta_k(\eta)$ are the time-dependent Bogoliubov coefficients. Here, we have introduced the positive ($+$) and negative ($-$) frequency modes
\begin{eqnarray}
    \phi_{k,\pm}(\eta)=\frac{e^{\mp i\int^{\eta}_{\eta_i} \omega_k(\tau)d\tau}}{\sqrt{2\omega_k(\eta)}}.\end{eqnarray}

\

Now, imposing that the modes satisfy   the condition
\begin{eqnarray}\label{zs_1}
\chi_{k}'(\eta)= -i\omega_k(\eta)\left(\alpha_k(\eta)\phi_{k,+}(\eta)-
\beta_k(\eta)\phi_{k,-}(\eta)\right),\end{eqnarray}
one can show that   the Bogoliubov coefficients must satisfy the system 
\begin{eqnarray}\label{alpha-beta}\left\{ \begin{array}{ccc}
\alpha_k'(\eta) &=& {\omega_k'(\eta)}\phi_{k,-}^2(\eta)\beta_k(\eta)\\
\beta_k'(\eta) &=& {\omega_k'(\eta)}\phi_{k,+}^2(\eta)\alpha_k(\eta),\end{array}\right.
\end{eqnarray}
in order for the  expression (\ref{zs}) to be a solution of the equation (\ref{KG}).

\

Finally, inserting (\ref{zs}) into the expression for the vacuum energy (\ref{vacuum-energy1}),
and taking into account that the Bogoliubov coefficients satisfy the equation $|\alpha_k(\eta)|^2- |\beta_k(\eta)|^2=1$,
one finds the following diagonalized form of the energy density \cite{Zeldovich}:
\begin{eqnarray}\label{vacuum-energy2}
\langle\rho(\eta)\rangle= \frac{1}{2\pi^2a^4(\eta)}\int_0^{\infty} k^2\omega_k(\eta)|\beta_k(\eta)|^2 dk,
\end{eqnarray}
where it is important to notice that $|\beta_k(\eta)|^2$ encodes the vacuum polarization effects and also the production of real particles, which are only produced when the adiabatic evolution breaks. In non-oscillating models this happens during the phase transition from the end of inflation to the beginning of kination, and fortunately the polarization effects disappear shortly after the beginning of kination, when the value of $|\beta_k(\eta)|$ stabilizes to a value which we will denote by $|\beta_k|$. Thus, it only encodes the production of real particles.

\

It is not difficult to show that the Bogoliubov coefficients stabilize, taking into account that during kination one has $a(t)\propto t^{1/3}$ and $H(t)\propto 1/t$, where $t$ is the cosmic time. Effectively,  
taking into account that the
modes that contribute to the particle production are those   which satisfy 
 $m_{eff}(\eta)a(\eta)\gg k$, we will have
\begin{eqnarray}
    \frac{\omega_k'(\eta)}{\omega_k(\eta)}=\frac{a'(\eta)a(\eta) m_{eff}^2(\eta)+a^2(\eta)m_{eff}'(\eta)m_{eff}(\eta)}
    {k^2+a^2(\eta)m_{eff}^2(\eta)}\sim \frac{a'(\eta)}{a(\eta)}=a(t)H(t)\propto t^{-2/3}.
    \end{eqnarray}

Therefore,  we conclude that the derivative of the Bogoliubov coefficients goes to zero,  meaning that they   stabilize. In fact, when the rate of expansion of the universe slows down,   the Bogoliubov coefficients always stabilizes,  because in that case,  $a(t)\propto t^{\frac{2}{3(1+w_{eff})}}$ where $w_{eff}$ denotes the effective Equation of State parameter, and thus
\begin{eqnarray}
    a(t)H(t)\propto t^{-\frac{1+3w_{eff}}{3(1+w_{eff})}},
\end{eqnarray}
which converges to zero when $w_{eff}>-1/3$, i.e., for a decelerating expansion.

\

Finally, we want to calculate 
\begin{eqnarray}\label{phi-average}
    \langle {\hat\phi}^2(\eta)\rangle=\frac{1}{4\pi^2a^2(\eta)}
    \int_0^{\infty}k^2\left(|\chi_k(\eta)|^2 -\frac{1}{2\omega_k(\eta)}\right)dk,
    \end{eqnarray}
where, as we have done with the energy density,   we have re-normalized it
subtracting the quantity $\frac{1}{2\omega_k(\eta)}$.

\

From the diagonalization method and using the system (\ref{alpha-beta}),  we can see that 
\begin{eqnarray}
    |\chi_k(\eta)|^2=\frac{|\beta_k(\eta)|^2}{\omega_k(\eta)}+    \frac{(|\beta_k(\eta)|^2)'}{\omega'_k(\eta)}+
    \frac{1}{2\omega_k(\eta)}\cong 
    \frac{|\beta_k|^2}{a(\eta)m_{eff}(\eta)}+    \frac{1}{2\omega_k(\eta)},   \end{eqnarray}
because the Bogoliubov coefficients stabilize during kination. Then,  inserting it in (\ref{phi-average})   we get
\begin{eqnarray}
    \langle {\hat\phi}^2(t)\rangle \cong \frac{1}{4\pi^2a^3(t)m_{eff}(t)}\int_0^{\infty}k^2|\beta_k|^2dk
    = \frac{ \langle N(t)\rangle}{2m_{eff}(t)}.
\end{eqnarray}


\begin{thebibliography}{99}

\bibitem{ellis}
J. Ellis, A. Linde, and D. Nanopoulos, Phys. Lett. {\bf B 118}, 59 (1982); D. Nanopoulos, K. Olive, and M. Srednicki, Phys.
Lett. {\bf B 127}, 30 (1983); J. Ellis, J. Kim, and D. Nanopoulos, Phys. Lett. {\bf B 145}, 181 (1984).


\bibitem{fkl0}
G. Felder, L. Kofman and  A. Linde, 
 {\it Instant Preheating},	Phys. Rev. {\bf D 59}, 123523 (1999)  	[arXiv:hep-ph/9812289]




\bibitem{fkl}
G. Felder, L. Kofman and  A. Linde,
{\it Inflation and preheating in NO models},
Phys. Rev. {\bf D 60}, 103505 (1999) [arXiv:hep-ph/9903350].


\bibitem{pv}
P. J. E. Peebles and  A. Vilenkin, {\it Quintessential Inflation},
Phys. Rev. {\bf D 59}, 063505 (1999)  [arXiv:astro-ph/9810509].

\bibitem{dimopoulos}
K. Dimopoulos, L.D. Wood and C. Owen, {\it Instant Preheating in Quintessential Inflation with $\alpha$-Attractors},  Phys. Rev. 
{\bf D 97}, 063525 (2018) [arXiv:1712.01760 [astro-ph.CO]].


\bibitem{Geng}
C. Q. Geng, C. C. Lee, M. Sami, E. N. Saridakis and A. A. Starobinsky, 
{\it Observational
constraints on successful model of quintessential Inflation}, 
JCAP {\bf 06}, 011 (2017) [arXiv:1705.01329 [gr-qc]].


\bibitem{campos}
A. H. Campos, H. C. Reis and  R. Rosenfeld,
{\it Preheating in Quintessential Inflation}, 
Phys. Lett. {\bf B 575}, 151-156 (2003)
	[arXiv:hep-ph/0210152].







\bibitem{haro23}
Jaume de Haro and Llibert Aresté Saló,
{\it Analytic formula to calculate the reheating temperature via gravitational particle production in smooth non-oscillating backgrounds},
Phys. Rev. {\bf D 107}, 063542 (2023)	[arXiv:2212.01276 [gr-qc]].



\bibitem{planck18}
Y. Akrami et al., 
{\it Planck 2018 results. X. Constraints on inflation}, 
A\&A {\bf 641}, A10 (2020)
[arXiv:1807.06211 [astro-ph.CO]].


\bibitem{haro21}
L. Aresté Saló, D. Benisty, E. I. Guendelman, J. de Haro,
{\it $\alpha$-attractors in Quintessential Inflation motivated by Supergravity}, Phys. Rev. {\bf D 103}, 123535 (2021)
	[arXiv:2103.07892 [astro-ph.CO]].
 	
{
\bibitem{linde}
Y. Akrami, R. Kallosh, A. Linde and  V. Vardanyan,
{\it Dark energy, $\alpha$-attractors, and large-scale structure 
surveys},
JCAP {\bf 1806}, 041 (2018) 	[arXiv:1712.09693 [hep-th]].
}


\bibitem{hossain}
Md. Wali Hossain, R. Myrzakulov, M. Sami and E. N. Saridakis,
{\it Unification of inflation and dark energy {\it à la} quintessential inflation},
Int. J. Mod. Phys. {\bf D24},  1530014 (2015) [arXiv:1410.6100 [gr-qc]].




{ \bibitem{Zeldovich}
Ya. B. Zeldovich and A. A. Starobinsky,
{\it Particle production and vacuum polarization in an anisotropic gravitational field}, JETP {\bf 34}, 1159 (1972).}

\bibitem{gmmbook}
A. A. Grib, S.G. Mamayev and V. M. Mostepanenko, {\it Vaccum Quantum Effects in Strong Fields},
Friedmann Laboratory Publishing for Theoretical Physics,
St. Petersburg (1994).




\end{thebibliography}
\end{document}